\documentstyle[aps,epsf,prb,preprint]{revtex}
\begin{document}
\title{Spin stiffness in the frustrated Heisenberg antiferromagnet}
\author{M.S.L. du Croo de Jongh, P.J.H. Denteneer } \address{
  Instituut-Lorentz, University of Leiden, P. O. Box 9506, 2300 RA
  Leiden, The Netherlands} \maketitle

\begin{abstract}
  We calculate the spin stiffness of the $S=\frac{1}{2}$ frustrated
  Heisenberg antiferromagnet directly from a general formula which is
  evaluated in the Schwinger boson mean-field approximation.  Both
  N\'eel and collinear ordering are considered.  For collinear
  ordering, we take the anisotropy of this phase into account, unlike
  previous approaches.  For N\'eel ordering, a detailed study is made
  of the finite-size scaling behavior of the two terms that make up
  the spin stiffness. The exponents of the scaling with the system
  size of the two terms comprising the spin stiffness turn out to be
  identical to those of the unfrustrated case.
\end{abstract}

\section{Introduction}

The recent interest in the frustrated Heisenberg antiferromagnets is
motivated by high $T_c$-superconductivity; the undoped compounds show
long-range antiferromagnetic order, similar to the Heisenberg model.
Upon doping superconductivity occurs. Adding frustration to the
Heisenberg model can be thought of as to mimic the effect of hole
doping.

We consider the frustrated Heisenberg model on a square lattice with
$N=L^2$ sites.  It is described by the following Hamiltonian for
quantum spins ${\bf S}_j$ on a lattice:
\begin{equation}
  H=J_1 \sum_{nn} {\bf S}_i \cdot {\bf S}_j +J_2 \sum_{nnn} {\bf S}_i
  \cdot {\bf S}_j ,\label{eq:heisenberg}
\end{equation}
where $nn$ denotes a pair $(ij)$ of nearest-neighbor sites and $nnn$ a
pair of next-nearest-neighbor sites. The spin length is fixed; $S =
\frac{1}{2}$.  Both $J_1$ and $J_2$ are taken to be non-negative. If
$J_2/J_1$ is small, the antiferromagnetic long-range order is
recovered (N\'eel-like).  For $J_2/J_1$ large, the system decomposes
in two N\'{e}el ordered sublattices which, however, have the same
quantization axis.  Alternating strips of up and down-spins will
occur; the so-called collinear ordering.  Clearly these couplings
frustrate each other.  If the spins were classical objects a large
number of phases, among which the N\'{e}el and collinear phases, would
become degenerate for $J_2/J_1 = 0.5$. For the quantum case, a quantum
phase transition to a spin-liquid phase might
occur.\cite{chandra,mila,ivanov,dots,einarsson,feiguin,schulz,sush,zhit}

In this paper, our intention is to employ the spin stiffness
$\rho_{\rm s}$ to measure magnetic order in the system. We calculate
the spin stiffness in the framework of the Schwinger boson mean-field
approximation (SBMFA) using a general formula for $\rho_{\rm s}$.
Previous evaluations of the spin stiffness were indirect
\cite{ivanov,auerbook,CHN}. However, apart from minor adjustments, we
confirm their results. Furthermore the scaling behavior of $\rho_s$ in
this approximation is derived. This is useful to sensibly extrapolate
results of more exact approaches, like quantum Monte Carlo and exact
diagonalization. Also, the typical system sizes for which scaling is
valid can be estimated in this way.\cite{einarsson,feiguin,schulz}

\section{Schwinger boson mean-field approximation}

The SBMFA improves upon standard mean-field theory by incorporating
correlations between neighboring spins \cite{ArAu,auerbook}. With this
approximation we derive the energies and wavefunctions of all states
of the frustrated Heisenberg model for both N\'{e}el and collinear
order. Our notation below will generally follow that of Mila et
al.\cite{mila}

The Schwinger boson transformation is a representation of the separate
spin operators by pairs of boson operators; $S^+=a^{\dagger}b$, $S^-=a
b^{\dagger}$, $ S^z={\scriptstyle \frac{1}{2}} \left[a^{\dagger} a -
b^{\dagger} b\right]$, supplemented by the local constraint
$a^{\dagger} a + b^{\dagger} b = 2S (=1)$.

\begin{figure}
  \centering \epsfxsize=8.6cm \epsffile{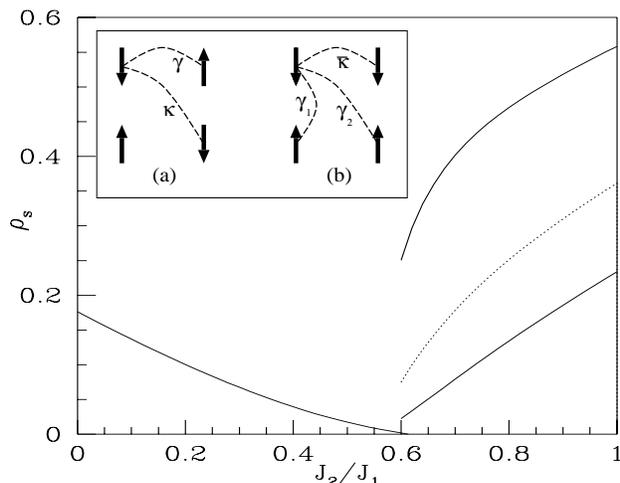}
\caption{\label{fig:rho_s} The spin stiffness $\rho_s$ in units of 
  $J_1$ (solid lines). For the collinear ordering the spin stiffness
  in the direction of the parallel spins, $\phi=0$, (lower solid
  curve) and in the direction of the antiparallel spins $\phi=\pi/2$
  (upper solid curve) are drawn. The dotted line is the result found
  by Ivanov and Ivanov \protect\cite{ivanov}. {\it Inset :} The mean
  fields for the N\'{e}el (a) and the collinear order (b). Mind
  $\kappa \sim D_{ij}$ and $\gamma \sim B_{ij}$.}
\end{figure}

To transform the Hamiltonian in a convenient form an appropriate
rotation in spin space is applied. Define $D_{ij} = a_i a_j^\dagger +
b_i b_j ^\dagger$ and $ B_{ij} = a_i b_j + b_i a_j$. The Hamiltonian
becomes
\begin{equation}
  H= -{\scriptstyle \frac{1}{2}} \sum_{AFM} J_{ij}(B_{ij}^\dagger
  B_{ij} -{\scriptstyle \frac{1}{2}}) + {\scriptstyle \frac{1}{2}}
  \sum_{FM} J_{ij} (D_{ij}^\dagger D_{ij} - {\scriptstyle
    \frac{3}{2}}), \label{eq:SB-H}
\end{equation}
where we have inserted $D_{ij}$ for the pairs $(ij)$ of spins parallel
in the $S^z$-direction and $B_{ij}$ for the antiparallel pairs.  For
the two orderings considered this is depicted in Figure
\ref{fig:rho_s}. The parameter $J_{ij}$ equals $J_1$ for
nearest-neighbors and $J_2$ for next-nearest-neighbors. The mean-field
decoupling is made using the fields $\kappa_{ij} = \frac{1}{2} \langle
D_{ij} \rangle $ and $ \gamma_{ij} =\frac{1}{2} \langle B_{ij}
\rangle$.  The local constraint $a^{\dagger} a + b^{\dagger} b = 1$ is
replaced by a global one and enforced by means of a Lagrange
multiplier $\lambda$. After a Fourier transform from $(a_i,b_i)$ to
$(a_{\bf p},b_{\bf p})$ and a Bogoliubov transformation from $(a_{\bf
  p},b_{\bf p})$ to $(\alpha_{\bf p},\beta_{\bf p})$;
\begin{eqnarray}
  a_{\bf p}&=& \alpha_{\bf p} \cosh \theta_{\bf p} + \beta_{\bf
    p}^\dagger \sinh \theta_{\bf p}, \nonumber \\ b_{-\bf p}^\dagger
  &=& \alpha_{\bf p} \sinh \theta_{\bf p} + \beta_{\bf p}^\dagger
  \cosh \theta_{\bf p} ,\nonumber \\ \tanh{2 \theta_{\bf p}} &=&
  \frac{\Delta_{\bf p}}{h_{\bf p} + \lambda}. \label{eq:bogoliubov},
\end{eqnarray}
the Hamiltonian becomes
\begin{equation}
  H_{MF}=E_c+ \sum_{\bf p} \omega_{\bf p} ( \alpha_{\bf p}^\dagger
  \alpha_{\bf p} + \beta_{\bf p} \beta_{\bf p}^\dagger),
  \label{eq:SBMFH2}
\end{equation}
where we have introduced the quasiparticle energies
\begin{equation}
  \omega_{\bf p}=\sqrt{(h_{\bf p}+\lambda)^2 - \Delta_{\bf p}^2}.
  \label{eq:omegaq}
\end{equation}
Below the quantities $h_{\bf p}$, $\Delta_{\bf p}$ and $E_c$ will be
defined for the N\'{e}el and collinear orderings separately.  The
fields $\kappa_{ij}$, $\gamma_{ij}$ and the Lagrange multiplier
$\lambda$ are obtained through consistency equations.

We consider two types of order:

{\it N\'{e}el order:} the specific form of $h_{\bf p}$, $\Delta_{\bf
  p}$, $E_c$ and the three consistency equations are:
\begin{eqnarray}
  h_{\bf p} &=& 4 J_2 \kappa \cos p_x \cos p_y ,\label{eq:hq} \\ 
  \Delta_{\bf p} &=& 2 J_1 \gamma (\cos p_x + \cos p_y)
  ,\label{eq:dq}\\ E_c &=& 2N\left [ J_1({\scriptstyle \frac{1}{4}} +
  2 \gamma^2) - J_2({\scriptstyle \frac{3}{4}} + 2 \kappa^2 ) -
  \lambda \right ], \\ \kappa &=& \frac{1}{N} \sum_{\bf p}
  \frac{h_{\bf p}+\lambda}{2 \omega_{\bf p}} \cos p_x \cos p_y
  \label{eq:sumkappa}, \\ \gamma &=& \frac{1}{N} \sum_{\bf p}
  \frac{\Delta_{\bf p}}{4 \omega_{\bf p}} (\cos p_x + \cos p_y)
  \label{eq:sumgamma}, \\ 1 &=& \frac{1}{N} \sum_{\bf p} \frac{h_{\bf
      p}+\lambda}{2 \omega_{\bf p}} .\label{eq:sum1}
\end{eqnarray}
For future applications it is also useful to define the ``condensate''
$m_s$ \cite{hirschtang} by
\begin{equation}
  m_s = \frac{h_{\bf 0} + \lambda}{ N \omega_{\bf 0}} = 1 -
  \frac{1}{N} \sum_{{\bf p} \neq (0,0),(\pi,\pi)} \frac{h_{\bf
      p}+\lambda}{2 \omega_{\bf p}}
  \label{eq:condensate}
\end{equation}
This is the combination of the -equivalent- ${\bf p}=(0,0)$ and ${\bf
  p}=(\pi,\pi)$ terms in (\ref{eq:sumkappa}) and (\ref{eq:sum1}),
which both diverge for $N \rightarrow \infty$. In the same limit, $N
\rightarrow \infty$, $m_s$ also equals the corresponding terms in
(\ref{eq:sumgamma}).

{\it Collinear order:} for this phase we introduce quantities with a
bar where confusion might arise;
\begin{eqnarray}
  \bar{h}_{\bf p} &=& 2J_1 \bar{\kappa} \cos p_x, \nonumber \\ 
  \bar{\Delta}_{\bf p} &=& 2 J_1 \gamma_1 \cos p_y + 4 J_2 \gamma_2
  \cos p_x \cos p_y ,\\ \bar{E}_c &=& 2 N\left ( J_1 (\gamma_1^2 -
  \bar{\kappa}^2-{\scriptstyle \frac{1}{4}}) + J_2 ({\scriptstyle
    \frac{1}{4}} + 2 \gamma_2^2) - \bar{\lambda} \right), \\ 
  \bar{\kappa} &=& \frac{1}{N} \sum_{\bf p} \frac{\bar{h}_{\bf
      p}+\bar{\lambda}}{2 \bar{\omega}_{\bf p}} \cos p_x,
  \label{eq:barkappa}\\ \gamma_1 &=& \frac{1}{N} \sum_{\bf p}
  \frac{\bar{\Delta}_{\bf p}}{2 \bar{\omega}_{\bf p}} \cos p_y,
  \label{eq:bargamma1}\\ \gamma_2 &=& \frac{1}{N} \sum_{\bf p}
  \frac{\bar{\Delta}_{\bf p}}{2 \bar{\omega}_{\bf p}} \cos p_x \cos
  p_y, \label{eq:bargamma2} \\ 1 &=& \frac{1}{N} \sum_{\bf p}
  \frac{\bar{h}_{\bf p}+\bar{\lambda}}{2 \bar{\omega}_{\bf p}}.
  \label{eq:bar1}
\end{eqnarray}
The condensate $\bar{m}_s$ and $\bar{\omega}_{\bf p}$ are defined in a
similar manner as before in (\ref{eq:condensate}) and
(\ref{eq:omegaq}).

The discussion above has given us the ground states $|0 \rangle$ with
energy $E_0=E_c+\sum_{\bf p} \omega_{\bf p}$ for both orderings. These
ground states are characterized by the absence of quasi-particles;
$\alpha|0 \rangle= \beta |0 \rangle=0$. Excited states are given by:
\begin{eqnarray}
  |a\rangle &=& \alpha_1^{\dagger}\cdot \dots \cdot \alpha_n^\dagger
  \cdot \beta_1^\dagger \cdot \dots \cdot \beta_m^\dagger |0\rangle,
  \label{eq:excitation}\\ H_{MF} |a \rangle &=& E_a|a \rangle \nonumber
  \\ & =&
  [E_0+\omega_1+\dots+\omega_n+\omega_1+\dots+\omega_m]|a \rangle.
  \label{eq:excitationE}
\end{eqnarray}

\section{The spin stiffness in the SBMFA} 

The spin stiffness $\rho_s$ is non-zero if there exists magnetic order
in the system and is, at $T=0$, associated with an increase in energy
upon twisting the order parameter of the system ($\Delta E=
\frac{1}{2} N \rho_s |{\bf q}|^2$ with ${\bf q}$ the wave vector of
the twist.)  \cite{leeuwen,fisher}. In line with Einarsson and Schulz
\cite{einarsson} we introduce this twist through a replacement of
$D_{ij}$ and $B_{ij}$ in (\ref{eq:SB-H}) by
\begin{eqnarray}
  D_{ij}({\bf q}) &=& a_i a_j^\dagger e^{\frac{i}{2} {\bf q} \cdot
    {\bf r}_\delta} + b_i b_j^\dagger e^{-\frac{i}{2} {\bf q} \cdot
    {\bf r}_\delta} \\ B_{ij}({\bf q}) &=& a_i b_j e^{\frac{i}{2} {\bf
      q} \cdot {\bf r}_\delta} + b_i a_j e^{-\frac{i}{2} {\bf q} \cdot
    {\bf r}_\delta}, \label{eq:dqbq}
\end{eqnarray}
with ${\bf r}_{\delta}={\bf r}_j-{\bf r}_i$. The resulting Hamiltonian
$H({\bf q})$ is now evaluated within the SBMFA.

Defining $\kappa_{ij}({\bf q}) = \frac{1}{2} \langle D_{ij}({\bf q})
\rangle$ and $\gamma_{ij}({\bf q}) = \frac{1}{2} \langle B_{ij}({\bf
  q}) \rangle$, the mean-field Hamiltonian becomes
\begin{eqnarray}
  H_{MF}({\bf q}) &=& - \sum_{AFM} J_{ij} \gamma_{ij}({\bf
    q})(B_{ij}^\dagger({\bf q})+B_{ij}({\bf q})-2 \gamma_{ij}({\bf
    q})] \nonumber \\ & &+ \sum_{FM} J_{ij} \kappa_{ij}({\bf
    q})(D_{ij}^\dagger ({\bf q}) + D_{ij}({\bf q}) - 2
  \kappa_{ij}({\bf q})] \nonumber \\ & & + \lambda \sum_{i}
  (a_i^\dagger a_i + b_i^\dagger b_i - 1)+\hbox{constants}.
\end{eqnarray}
Since within the SBMFA we know what the excited states are [see
(\ref{eq:excitation}) and (\ref{eq:excitationE})], the spin stiffness
can be directly evaluated from second order perturbation theory
\begin{equation}
  \rho_s= -\frac{1}{N}\langle 0|t|0\rangle + \frac{2}{N}\sum_{|a
    \rangle \neq |0 \rangle} \frac{|\langle 0|j|a\rangle|^2}{E_0-E_a}
  \equiv T+J,
  \label{eq:rhofinal}
\end{equation}
with the quantities $t$ and $j$ defined by
\begin{equation}
  t= \left.- \frac{d^2}{dq^2} H_{MF}({\bf q}) \right |_{{\bf q}=0} ,~
  j = \left.  \frac{d}{dq} H_{MF}({\bf q}) \right |_{{\bf q}=0}.
  \label{eq:tjdef}
\end{equation}
In (\ref{eq:rhofinal}), we have also defined the abbreviations $T$ and
$J$ for the two terms in $\rho_s$.

\section{Results for the spin stiffness}

From this point on, we set ${\bf q} = q ( \cos \phi, \sin \phi )$.  Of
the two terms for $\rho_s$ in (\ref{eq:rhofinal}), $T$ is evaluated
more easily. We obtain after straightforward manipulations with
Brillouin zone summations :
\begin{eqnarray}
  T &=& J_1 \gamma^2 - 2 J_2 \kappa^2 ,\label{eq:TYneel}\\ \bar{T} &=&
  2 J_2 \gamma^2_2 + J_1(\gamma_1^2 \sin^2 \phi - \bar{\kappa}^2
  \cos^2 \phi ), \label{eq:TYcoll}
\end{eqnarray}
These simple equations hold for all system sizes $N$.

The quantity $J$ requires more effort; the operator $j$ has to be
expressed in the operators $\alpha_{\bf p}$ and $\beta_{\bf p}$
defined in (\ref{eq:bogoliubov}). For the wavefunctions $|a \rangle$
and energies $E_a$ of the excitations we use (\ref{eq:excitation}) and
(\ref{eq:excitationE}).  The resulting values for $J$ are written as
summations over the Brillouin zone:
\begin{eqnarray}  
  J &=& - \frac{1}{N} \sum_{\bf p} \frac{\sin^2 p_y}{\omega_{\bf p}^3}
  ( J_1 \gamma (h_{\bf p}+\lambda) - 2 J_2 \kappa \Delta_{\bf p} \cos
  p_x )^2 \nonumber, \\ \bar{J} &=& -\frac{1}{N} \sum_{\bf p}
  \frac{1}{\bar{\omega}_{\bf p}^3} ( \cos \phi \sin p_x [J_1
  \bar{\kappa} \bar{\Delta}_{\bf p} - 2 J_2 \gamma_2(\bar{h}_{\bf p} +
  \bar{\lambda}) \cos p_y] \nonumber \\ & & - \sin \phi \sin p_y[J_1
  \gamma_1(\bar{h}_{\bf p} + \bar{\lambda}) + 2 J_2 \gamma_2
  (\bar{h}_{\bf p} +\bar{\lambda})\cos p_x] )^2. \nonumber
\end{eqnarray}
Only for the infinitely large lattice these equations can be
simplified by replacing summations by integrals and partially
integrating. The expression for the spin stiffness $\rho_s=T+J$ then
simplifies considerably and becomes:
\begin{eqnarray}
  \rho_s &=& m_s(J_1 \gamma - 2 J_2 \kappa),
  \label{eq:rhosneel}\\ \bar{\rho}_s &=& \bar{m}_s [ (2 J_2 \gamma_2-
  J_1 \bar{\kappa})\cos^2 \phi \nonumber \\ & & + (2J_2 \gamma_2 + J_1
  \gamma_1)\sin^2 \phi ] \label{eq:rhoscoll}.
\end{eqnarray}

Ivanov and Ivanov \cite{ivanov} apply a different method to derive
$\rho_s$. They use a modified spin wave theory which leads to the same
consistency equations as the Schwinger boson approach. $\rho_s$ is
then obtained by calculating the correlation length $\xi$ associated
with the spin-spin correlation function $\langle {\bf S}_i \cdot {\bf
  S}_j \rangle$ and comparing this $\xi$ to the expression for $\xi$
obtained for the non-linear sigma model to two-loop order by
Chakravarty et al. \cite{CHN} (where $\xi \sim \exp(2\pi
\rho_s/\theta)$ with $\theta$ the temperature).  For N\'{e}el ordering
their expression is identical to our result (\ref{eq:rhosneel}).  It
is gratifying to see that the non-linear sigma model also is the
effective field theory for the low-energy physics of the {\it
  frustrated} Heisenberg antiferromagnet.

For the collinear ordering they obtain the geometrical average of our
$\cos^2 \phi$ and $\sin^2 \phi$-terms, whereas we take explicitly the
anisotropy of this phase into account, Still both expressions for
$\bar{\rho}_s$ vanish at the same value of $J_2/J_1$.

Table \ref{tab:TJ} and Figure \ref{fig:rho_s} contain our numerical
results.

\section{Scaling of the spin stiffness}

It is necessary to know the size dependence of observables to obtain a
good approximation for their limit values. Neuberger and Ziman
\cite{neuberger} derived the scaling behavior for an unfrustrated
Heisenberg antiferromagnet explicitly.  Here we extend this to the
case of frustration.

Recently some discussion has arisen about where the scaling behavior
of $\rho_s$ sets in\cite{feiguin}. Our formulas in the last section
lend themselves well to investigate this.

Here we only treat the N\'{e}el ordering. We want to know the scaling
behavior of the condensate $m_s$ and the two terms $J$ and $T$ that
make up $\rho_s$ ($\rho_s=T+J$). The latter two will turn out to have
different scaling behavior.

As can be seen from (\ref{eq:TYneel}) only the scaling behavior of
$\kappa_N$ and $\gamma_N$ is required for $T$. These two are part of
the set $(\kappa_N,\gamma_N,\lambda_N)$ of mutually dependent
quantities. We will now argue what is the exponent of their scaling
behavior and therefore of $T$, without trying to obtain the precise
prefactor (which would be quite tedious).

Name the equations (\ref{eq:sumkappa})-(\ref{eq:sum1}) $I$, $II$, and
$III$, respectively. They contain poles at ${\bf p}=(0,0)$ and ${\bf
  p}=(\pi,\pi)$. With help of (\ref{eq:omegaq}), (\ref{eq:hq}) and
(\ref{eq:dq}) we rearrange them as $I-III$, $II-III$ and $4 J_2
\kappa_N I - 4 J_1 \gamma_N II + \lambda_N III$. We neglect the ${\bf
  p}=(0,0)$ and ${\bf p}=(\pi,\pi)$ terms in the summations. It is
easy to show that this will give rise to errors of the order
$O(N^{-2})$. Next we expand these equations to first order round their
infinite size values $(\kappa,\gamma,\lambda)$. Define the size
dependences $\delta \kappa_N=\kappa_N-\kappa$, $\delta \gamma_N =
\gamma_N - \gamma$ and $\delta \lambda_N= \lambda_N-\lambda$ to obtain
the equation
\begin{equation}
  \left [\begin{array}{c} \kappa-1 \\ \gamma-1 \\ \lambda -4 J_1
    \gamma^2+ 4 J_2 \kappa^2 \end{array} \right ] = \vec{A}
    +\vec{\vec{B}} \cdot \left [ \begin{array}{c} \delta \kappa_N \\ 
    \delta \gamma_N \\ \delta \lambda_N \end{array} \right ],
    \label{eq:roughly}
\end{equation}
where $\vec{A}$ and $\vec{\vec{B}}$ contain summations over the
Brillouin zone dependent on the infinite size parameters $\kappa$,
$\gamma$, and $\lambda$.  The remaining size dependence of
$\vec{\vec{B}}$ can be neglected as it leads to higher order terms. On
the other hand the summations in $\vec{A}$ will be replaced by
integrations plus size dependent corrections. Using Neuberger and
Ziman \cite{neuberger} we obtain
\begin{equation}
  \vec{\vec{B}}\cdot \left [ \begin{array}{c} \delta \kappa_N \\ 
  \delta \gamma_N \\ \delta \lambda_N \end{array} \right ] =
  \frac{1}{N^{3/2}} \vec{C}.
\end{equation}
The parameters $\kappa_N$, $\gamma_N$ and $\lambda_N$ thus scale with
$N^{-3/2}$. A direct consequence of this is that $T_N-T \sim
N^{-3/2}$. If the size dependence of the parameters is neglected
[$(\kappa_N,\gamma_N,\lambda_N) \rightarrow (\kappa,\gamma,\lambda)$
{\it inside} the summations (\ref{eq:sumkappa})-(\ref{eq:sum1})], we
also find the prefactor:
\begin{equation}
  T_N-T = \frac{0.7186}{N^{3/2}} \frac{1}{\sqrt{2}}
  \sqrt{\frac{\lambda +4J_2 \kappa}{\lambda - 4 J_2 \kappa}} \left[
  \lambda - 8 J_2 \kappa \right ]. \label{eq:Tscalesimple}
\end{equation}
Upto $J_2/J_1 \approx 0.5$ this is in good agreement with the
numerical solution of (\ref{eq:sumkappa})-(\ref{eq:sum1}) for various
sizes $N$.

\begin{figure}
  \centering \epsfxsize=8.6cm \epsffile{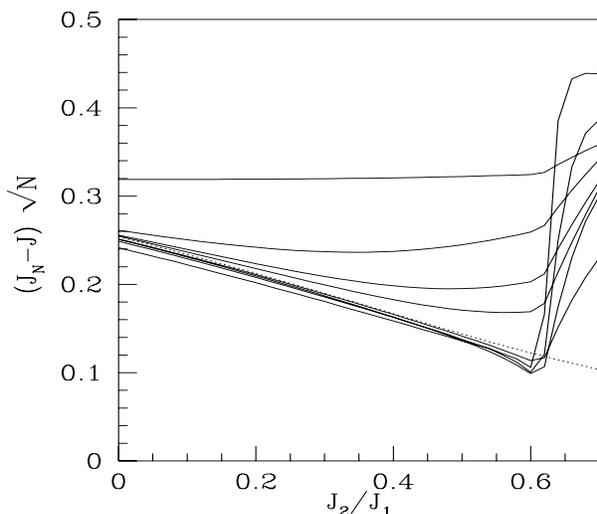}
  \caption{\label{fig:JYscale} The numerical scaling behavior of
    $J_N$ (solid lines) for sizes $L=2,4,6,8,10,20,40,100$ (numbering
    is top-down) compared with the theoretical formula
    (\ref{eq:Jscale}) (dotted line).}
\end{figure}

Next we consider the condensate $m_{s,N}$ and $J_N$, for which we can
even derive the prefactors.  Replacement of
$(\kappa_N,\gamma_N,\lambda_N)$ by $(\kappa,\gamma,\lambda)$ in the
summation of (\ref{eq:condensate}) leads to errors of the order
$O(N^{-3/2})$. We will neglect these.  Neuberger and Ziman
\cite{neuberger} state a lemma applicable to this summation. This
leads to:
\begin{equation}
  m_{s,N} - m_s = \frac{0.6208}{\sqrt{N}} \sqrt{2} \sqrt{\frac{\lambda
      + 4 J_2 \kappa}{\lambda - 4 J_2 \kappa}}. \label{eq:Sscale}
\end{equation}
For $J_N$ we find in a similar fashion:
\begin{equation}
  J_N-J = \frac{0.6208}{\sqrt{N}} \frac { \sqrt{\lambda^2 - (4 J_2
      \kappa)^2}}{4 \sqrt{2}}.\label{eq:Jscale}
\end{equation}
These formulas are in excellent agreement with the numerical solution
to (\ref{eq:sumkappa})-(\ref{eq:sum1}). For $J_N$ this is depicted in
figure \ref{fig:JYscale}.

In conclusion, we have obtained the scaling behavior of $T$ and $J$
from an analysis of the formulas in the SBMFA for the frustrated
Heisenberg antiferromagnet. The qualitative scaling behavior (i.e.
the exponents) is the same as for the {\it unfrustrated} case, which
was discussed by Neuberger and Ziman \cite{neuberger}. The scaling
behavior that was utilized by Einarsson and Schulz \cite{einarsson}
does agree with our findings but we confirm the message of Feiguin et
al.\cite{feiguin} that the clusters they used are too small for the
scaling behavior of $J$ to have set in.  Numerically we see that the
scaling behavior starts around size $N=100$ (See Figure
\ref{fig:JYscale}) whereas the largest cluster they used is $N=36$.

We would like to thank J.M.J. van Leeuwen and W. van Saarloos for
fruitful discussions.

\begin{table}
  \centering
\begin{tabular}{||c|c|cc|cc|cc||}
  $J_2/J_1$ & Order &$T$ & &$-J$ & & $\rho_s$ & \\ \hline 0.0 & N &
  0.3352 & & 0.1596 & & 0.1757 & \\ 0.1 & N & 0.2961 & & 0.1596 & &
  0.1365 & \\ 0.2 & N & 0.2597 & & 0.1597 & & 0.1000 & \\ 0.3 & N &
  0.2271 & & 0.1600 & & 0.0672 & \\ 0.4 & N & 0.1995 & & 0.1604 & &
  0.0391 & \\ 0.5 & N & 0.1783 & & 0.1612 & & 0.0171 & \\ 0.6 & N &
  0.1639 & & 0.1622 & & 0.0017 & \\ \hline 0.6 & C & 0.6231 & 0.0499 &
  0.3841 & 0.0284 & 0.2390 & 0.0214 \\ 0.7 & C & 0.7211 & 0.1563 &
  0.3327 & 0.0797 & 0.3884 & 0.0766 \\ 0.8 & C & 0.7794 & 0.2566 &
  0.3232 & 0.1264 & 0.4563 & 0.1302 \\ 0.9 & C & 0.8327 & 0.3492 &
  0.3292 & 0.1689 & 0.5034 & 0.1803 \\ 1.0 & C & 0.8866 & 0.4361 &
  0.3434 & 0.2086 & 0.5432 & 0.2275 \\ 1.1 & C & 0.9420 & 0.5190 &
  0.3623 & 0.2465 & 0.5797 & 0.2725 \\ 
\end{tabular}
\caption{\label{tab:TJ} The limit ($N=\infty$) values for $T$, $-J$ and
  $\rho_s$ as function of the ratio $J_2/J_1$ where $J_1=1$. The two
  orderings considered are N\'eel (N) and collinear (C) order. For the
  collinear ordering there are two directions: along the antiparallel
  spins, $\phi=\pi/2$ (listed first) and along the parallel spins,
  $\phi=0$ (listed secondly).}
\end{table}

\begin{references}
\bibitem{chandra} P. Chandra, B. Doucot, Phys. Rev. B {\bf 38}, 9335
  (1988).  \bibitem{mila} F. Mila, D. Poilblanc, C. Bruder, Phys. Rev.
  B {\bf 43}, 7891 (1991).  \bibitem{ivanov} N.B. Ivanov, P. Ch.
  Ivanov, Phys. Rev. B {\bf 46}, 8206 (1992).  \bibitem{dots} A.V.
  Dotsenko, O.P. Sushkov, Phys. Rev. B {\bf 50}, 13821 (1994).
\bibitem{einarsson} T. Einarsson, H.J. Schulz, Phys. Rev. B {\bf 51},
  6151 (1995).  \bibitem{feiguin} A.E. Feiguin, G.J. Gazza, A.E.
  Trumper, H.A. Ceccatto, Phys. Rev. B {\bf 52}, 15043 (1995).
\bibitem{schulz} H.J. Schulz, T.A.L. Ziman, D. Poilblanc, J. Phys. 1
  France {\bf 6}, 675 (1996).  \bibitem{sush} O.P. Sushkov, cond-mat
  9602150.
 \bibitem{zhit} M. Zhitomirsky, K. Ueda, Phys. Rev. B {\bf 54}, 9007 (1996)
%
 \bibitem{ArAu} D.P. Arovas, A. Auerbach, Phys. Rev. B {\bf 38}, 316
   (1988), A. Auerbach, D.P. Arovas, Phys. Rev. Lett. {\bf 61}, 617
   (1988).  \bibitem{auerbook} A. Auerbach, {\em Interacting Electrons
     and Quantum Magnetism} (Springer, New York, 1994).  \bibitem{CHN}
   S. Chakravarty, B.I.  Halperin, D.R.  Nelson, Phys. Rev. B {\bf
     39}, 2344 (1988).  \bibitem{hirschtang} J.E. Hirsch, S. Tang,
   Phys. Rev. B {\bf 39}, 2850
   (1989).
%

 \bibitem{fisher} M.E. Fisher, M.N. Barber, D. Jasnow, Phys. Rev. A
   {\bf 188}, 1111 (1973).  \bibitem{leeuwen} J.M.J. van Leeuwen,
   M.S.L. du Croo de Jongh, P.J.H.  Denteneer, J. Phys. A:Math. Gen.
   {\bf 29}, 41 (1996).  \bibitem{neuberger} H.  Neuberger, T. Ziman,
   Phys. Rev. B {\bf 39}, 2608 (1989).
\end{references}
\end{document}